\documentclass[10pt]{article}
\usepackage{amssymb}
\usepackage{amsmath}
\usepackage{graphicx}
\usepackage{cite}
\usepackage{float}
\usepackage{caption}
\usepackage{authblk}
\usepackage{subfigure}
\usepackage[margin=1.0in]{geometry}
\usepackage{color}

\begin{document}

\title{ Combination of equilibrium and non-equilibrium carrier statistics into an atomistic quantum transport model for tunneling hetero-junctions
}
\author[1*]{\underline{Tarek A. Ameen}}
\author[1*]{\underline{Hesameddin Ilatikhameneh}}
\author[1]{{Jun Z. Huang}}
\author[1]{Michael Povolotskyi}
\author[1]{Rajib Rahman}
\author[1]{Gerhard Klimeck}
\affil[1]{\normalsize{Network for Computational Nanotechnology, Department of Electrical and Computer Engineering, Purdue University, West Lafayette, IN 47907, USA}}
\affil[*]{\normalsize{These authors contributed equally to this work.}}
\renewcommand\Authands{ and }
\maketitle
\providecommand{\keywords}[1]{\textbf{\textit{Keywords---}} #1}
\maketitle

\begin{abstract}


Tunneling hetero-junctions (THJs) usually induce confined states at the regions close to the tunnel junction which significantly affect their transport properties. Accurate numerical modeling of such effects requires combining the non-equilibrium coherent quantum transport through tunnel junction, as well as the quasi-equilibrium statistics arising from the strong scattering in the induced quantum wells. In this work, a novel atomistic model is proposed to include both effects: the strong scattering in the regions around THJ and the coherent tunneling. The new model matches reasonably well with experimental measurements of Nitride THJ and provides an efficient engineering tool for performance prediction and design of THJ based devices.


\end{abstract}

\section{Introduction}
\begin{figure}[H]
\centering
\includegraphics[width=60mm]{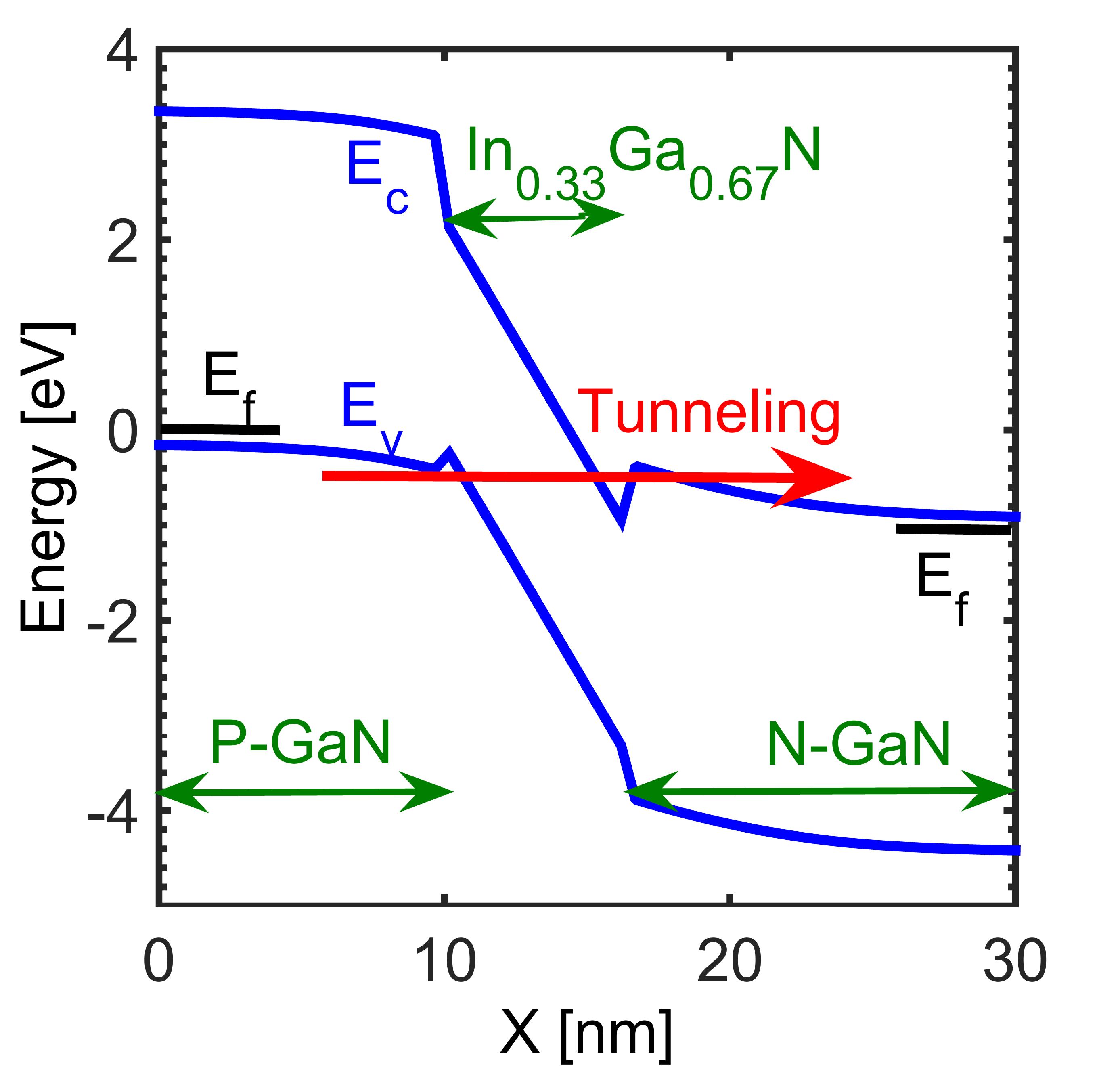}
\caption{Energy band diagram of a tunneling hetero-junction. The band diagram corresponds to the InGaN/GaN tunneling diode reported in \cite{krishnamoorthy2010polarization}.}
\label{Band_diagram_fig}
\end{figure}

Tunneling hetero-junctions (THJs) are of great importance to many electronic and opto-electronic applications. High performance tunnel transistors\cite{Wenjun1,dewey2011fabrication}, multi-junction solar cells\cite{bedair2016high}, resonant tunneling diodes \cite{bowen1997quantitative}, and light emitting diodes \cite{takeuchi2001gan} have all been improved by using THJs. A THJ is composed of heavily doped P and N regions separated by another material with a smaller band gap ($E_g$), as shown in Figure (\ref{Band_diagram_fig}). The combination of a large $E_g$ substrate material and a small $E_g$ tunneling quantum well is desired in all of these mentioned applications, although for different reasons. This paper is devoted to introducing a novel atomistic quantum transport model developed to simulate and design THJ based devices.  

The challenge in developing an accurate model for THJs is that coherent quantum transport in the tunneling region (Non-equilibrium) needs to be combined with quasi-equilibrium statistics arising from the strong scattering in the adjacent heavily doped regions. This multi-physics problem involves many physical processes that need to be captured in this model properly and yet efficiently to be used as a predictive tool. These processes are: (1) quantum confinement effects, (2) band to band tunneling, (3) heavy N and P doping and large charge densities that give rise to strong electron-electron (e-e) scattering resulting in quasi-equilibrium, (4) scattering induced broadening and filling of confined states in the quantum well, (5) strain induced deformation of band structure, (6) internal piezo- and pyro-electric polarization in wurtzite materials (i.e. III-Nitrides).
    
A quantum transport method is essential to account for confinement effects and tunneling processes properly, but including the strong e-e scattering in the heavily doped P and N regions is not trivial and is computationally unfeasible. Such strong scattering dominates the transport dynamics in the quantum well region and should be taken into account. Koswatta et al. have studied the impact of electron-phonon scattering on the  performance of THJs using one band quantum transport simulations\cite{koswatta2010possibility}. Later on, a full band atomistic electron-phonon scattering model considering the modified phonon spectra of a nanoscale device has been developed \cite{luisier2009atomistic}. Neverthless, it has been concluded that the electron-phonon scattering by itself does not completely thermalize the carriers in quantum wells \cite{koswatta2010possibility}. In this work, we propose an efficient model that accounts for the strong scattering in the heavily doped regions and coherent quantum transport in the tunneling region. Trap assisted tunneling can contribute to the current in low quality cases with high defect density. Such effect is beyond the scope of this paper and has been neglected.
 
The idea of combining thermalization and equilibrium statistics in non equilibrium Green's function (NEGF) was  used to simulate resonant tunneling diodes (RTDs) and supperlattices \cite{klimeck1995quantum,long2016performance} by partitioning the simulation domain to equilibrium (Eq) and non-equilibrium (Neq) regions. Such a coupled Eq-Neq quantum transport model was needed to explain experimental measurements of RTDs \cite{klimeck1995quantum}. But this model \cite{klimeck1995quantum} does not work for band to band tunneling in which the tunneling boundaries vary with energy and cannot be applied to THJs. Therefor a more sophisticated model is proposed here for THJs.
Current-voltage characteristics computed from this new model is shown to agree well with experimental measurements\cite{krishnamoorthy2010polarization} of Nitride THJs over ballistic simulations.

\section{A combined equilibrium - non equilibrium  model}
Non-equilibrium Green's function (NEGF) formalism is a well established and widely accepted quantum transport approach \cite{dattaold,dattanew}. NEGF can capture accurately the band to band tunneling (BTBT) in THJs. In addition to this, the inherent quantum mechanics in NEGF is needed for confinement effects that exist in the quantum well regions and, more generally, may arise from device geometry (i.e. in a nanowire or an ultra thin body device).
Given a certain potential distribution $V$ across the device, the retarded Green's function ($G^{R}$) and the spectral function ($A$) are given by \cite{dattaold,dattanew}
\begin{equation}
G^{R}(K_{\perp},E)=\left[EI-H(K_{\perp})+qV-\Sigma_s^R-\Sigma_d^R-\Sigma_{sc}^R\right]^{-1},
\label{eqn_1}
\end{equation}
\begin{equation}
A(K_{\perp},E)=i \left[G^{R}(K_{\perp},E)-G^{R}(K_{\perp},E)^{\dagger}\right],
\label{eqn_2}
\end{equation}
where $E$, $I$, $H$, and $K_{\perp}$ are the carrier energy, identity matrix, device Hamiltonian, and the wave vector associated with periodicity in the transverse direction respectively. $\Sigma_s^R$ and $\Sigma_d^R$ are self-energies due to open boundaries of source and drain lead respectively, while $\Sigma_{sc}^R$ is the self-energy due to different scattering mechanisms. All of the self-energies are functions of $(K_{\perp},E)$ in general. The diagonal elements of the spectral matrix ($A$) are the local density of states on individual atoms. The local density of states in the device is calculated quantum mechanically in the framework of a multi-band tight binding Hamiltonian. Depending on the nature of theses states, they are either filled coherently or thermally. States that exist in regions with strong scattering are filled with Fermi-Dirac statistics giving equilibrium free charge $n_{eq}(r)$. While states that exist in non equilibrium regions are filled coherently giving coherent free charge $n_{coh}(r)$. 
\begin{equation}
n_{eq}(r)=\int \frac{dK_{\perp}}{2\pi} \int \frac{dE}{2\pi}  A(K_{\perp},E)_{r,r} ~~ f_{FD}(E-E_{fr})
\label{eqn_3}
\end{equation}
where  $f_{FD}$ is the Fermi-Dirac distribution function and  $E_{fr}$ is the quasi-Fermi level at position $r$.
\begin{equation}
n_{coh}(r)=\int \frac{dK_{\perp}}{2\pi} \int \frac{dE}{2\pi}  \left[G^{R}(K_{\perp},E) (f_s\Gamma_s+f_d\Gamma_d) G^{R}(K_{\perp},E)^{\dagger}\right]_{r,r},
\label{eqn_4}
\end{equation}
where  $f_s$ and $f_d$ are Fermi-Dirac occupations of $(K_{\perp},E)$  in source and drain, $\Gamma_s$ and $\Gamma_d$ are source and drain broadening matrices $\Gamma=i(\Sigma^R-\Sigma^{R\dagger})$.
These free charges are input to the 3D finite-element Poisson equation which reads as follows :
\begin{equation}
\overrightarrow{\nabla} \cdot \left(-\varepsilon \overrightarrow{\nabla} V + \overrightarrow{P}_{pyro }+ \overrightarrow{P}_{piezo } \right)= \rho, 
\label{eqn_20}
\end{equation}
where $V$, $\varepsilon$, $\rho$, $\overrightarrow{P}_{pyro }$, and $\overrightarrow{P}_{piezo }$ are the electrostatic potential, dielectric constant, total charge (ionized doping plus free charge densities $n_{eq}$,$n_{coh}$), Pyro- and Piezo-electric polarization fields respectively. The Poisson equation is solved self consistently with the Green's function.

The tunneling current is calculated in the non-equilibrium region using
\begin{equation}
I=\int \frac{dK_{\perp}}{2\pi} \int \frac{dE}{2\pi} T(K_{\perp},E) (f_s\Gamma_s-f_d\Gamma_d),
\label{eqn_5}
\end{equation}

where $T$ is the transmission\cite{dattanew}. There are two critical points that need to be uniquely treated for THJs; First, the boundary between what is considered equilibrium (or strong scattering) regions and what is considered non-equilibrium (or coherent) region. Second, an efficient way to include scattering in the Green's function.

\begin{figure}[H]
\centering
\includegraphics[width=60mm]{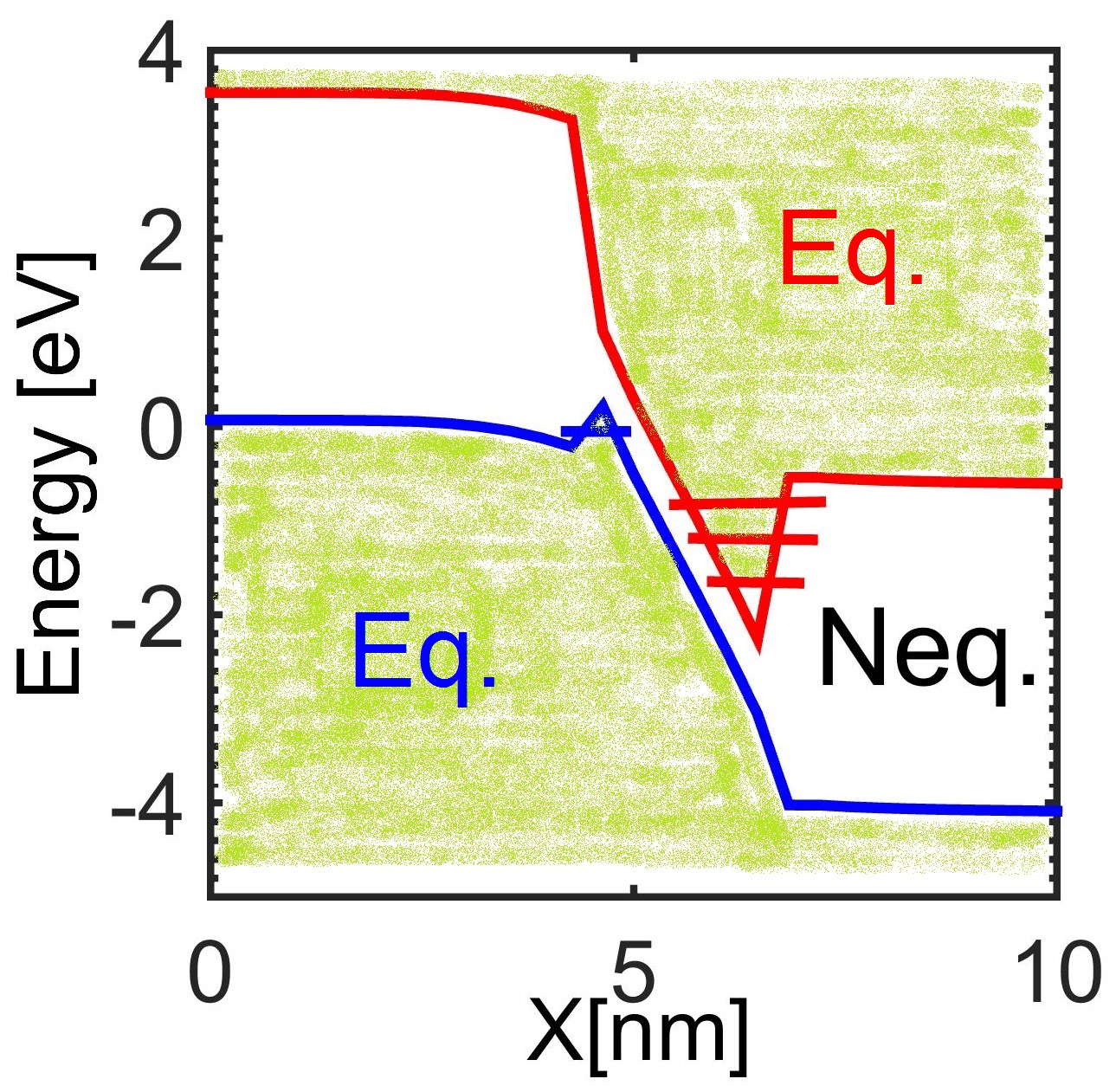}
\caption{A schematic diagram showing the boundaries between equilibrium (Eq.) and non-equilibrium (Neq.) regions at different energy points at a certain $K_{\perp}$.  The white region is Neq. domain and the green dotted regions are Eq. domains.}
\label{EQNEQ_fig}
\end{figure}

A reasonable boundary between equilibrium and non-equilibrium (Eq-Neq) regions can be deduced from band edges which are position and $K_{\perp}$ dependent. Fig. \ref{EQNEQ_fig} shows a  schematic diagram  of the boundaries between equilibrium and non-equilibrium (Eq-Neq) regions. Source valence states and drain conduction states have frequent scattering events and are considered to be in quasi-thermal equilibrium due to high carrier densities. While the band gap has low density of states and carriers can traverse it predominantly by the band to band tunneling process, therefore the transport across the gap is treated coherently (Neq). At each $K_{\perp}$, a band edge diagram as shown in Fig. \ref{EQNEQ_fig} is calculated along the transport direction, which is then used to determine the boundaries between equilibrium and non-equilibrium domains at each energy $E$.   

The second point is to include the inherent strong scattering in the equilibrium regions via the scattering self-energy $\Sigma_{sc}$. The scattering in the heavily doped regions has two significant effects on the tunneling process. First, it broadens the energies of the confined states in the quantum well. Second, occupation of those states by charge carriers, i.e. strong scattering leads to quasi-equilibrium which is assumed to result in a Fermi-Dirac distribution. The broadening and filling of the quantum well states induced by scattering play a significant role in determining how many carriers will tunnel from the P to the N region. The scattering can come from many sources; defects, alloy, impurity, phonons, and (e-e) interaction. Except for the latter, the scattering mechanisms can be included either explicitly in the Hamiltonian to account for defects, random alloy, and impurities, or it can be included as a perturbation self-energy in the case of electron-phonon interactions. Such scattering mechanisms complicate the model, require more computational resources, and additional self consistent Born loop between Green's function and the self-energies. The amount of resources, time and memory, required to include these scattering mechanisms in a formal way is quite significant and can be orders of magnitude more than the model we are proposing. In addition, the e-e scattering is not negligible due to the large density of charge arising from the heavy doping. Furthermore, inclusion of the e-e scattering is not trivial due to the many-body nature of these interactions. Moreover, there is no formal scattering self-energy that leads to full thermalization in high carrier density regions. Hence in this work, the scattering induced broadening $\eta$, $\Sigma_{sc}=i\eta$, of the states is added explicitly to the on-site Hamiltonian elements in the equilibrium regions with boundaries dependent on $E$ and $K_{\perp}$ as described earlier. Then the charge filling of the equilibrium states is calculated using Fermi-Dirac statistics as shown in equation (\ref{eqn_3}).  This broadening $\eta$ is related to the energy relaxation time with a Heisenberg uncertainty, $\eta ~ \tau \approx \frac{\hbar}{2}$. The energy relaxation time can be calculated from the carrier mobility $\mu$, as $\tau =\frac{m^* \mu}{q}$, where $m^*$ is the effective mass. Hence the broadening $\eta$ is given by
\begin{equation}
\eta \approx \frac{q \hbar}{2m^* \mu}.
\label{eqn_6}
\end{equation}
Electrons and holes have different mobilities and consequently different broadening $\eta_c$ and $\eta_v$. In general, the energy dependent scattering rate, and consequently the broadening $\eta$, is  a function of the density of states. For simplicity $\eta_c$ and $\eta_v$ are assumed to be energy independent  for energies beyond the confined band-edges, and to decay exponentially inside the band gap with the same factor as the density of states.
\begin{equation}
\eta_{gap} = \eta e^{-\frac{|E-E_{c/v}|}{\Delta}},
\label{eqn_7}
\end{equation}
where $\Delta$ is the band tail decay energy (Urbach energy). 
The model has been implemented in the Nanoelectronics Modeling tool NEMO5\cite{nemo5,nemo5_2,sellier2012nemo5,ilatikhameneh2016fowler}.

\section{Results and discussion}
In order to verify the validity of the model, we have simulated the PN junction reported in \cite{krishnamoorthy2010polarization}. This PN junction is composed of a 6.4nm In$_{0.33}$Ga$_{0.66}$N quantum well sandwiched within a GaN substrate. The P-GaN is Mg doped with a doping concentration of $N_A=10^{19}cm^{-3}$ and the N-GaN is Si doped with a concentration of $N_D=5\times 10^{18}cm^{-3}$.
The doping model considers incomplete ionization and the ionization energies of Mg and Si are about $0.17$meV and $0.133$meV respectively\cite{lee2007characteristics,nakamura1992thermal,irokawa2005electrical}. The Hamiltonian and self-energies in NEGF are constructed from sp$^3$ nearest neighbor tight-binding model\cite{schulz2006tight}. The 3D Poisson's equation is solved self consistently with NEGF. The equation for the InGaN/GaN polarization coefficient is in the Appendix.

\begin{figure}[H]
\centering
\includegraphics[width=80mm]{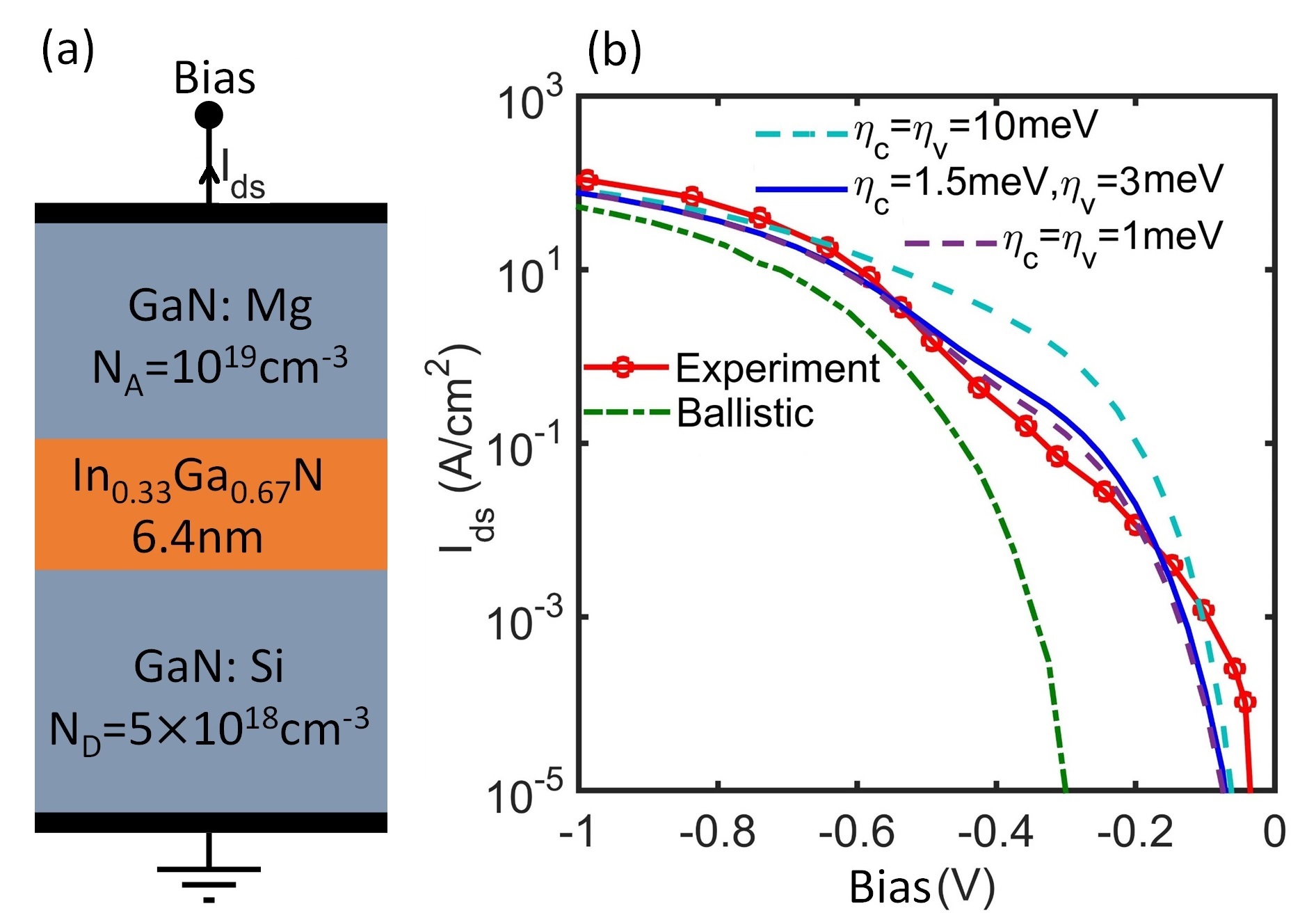}
\caption{a) A schematic diagram of the device measured in \cite{krishnamoorthy2010polarization}.  b) Computed current-voltage characteristics for the device of Ref.  \cite{krishnamoorthy2010polarization}. The scattering model reasonably matches the experimental measurements unlike the ballistic model that significantly underestimates the current.}
\label{IV_fig}
\end{figure}

For heavily doped GaN and InN, the Urbach band tail energies are $\Delta_{GaN} = 50$meV \cite{qiu1995study} and $\Delta_{InN} = 28$meV \cite{wu2003temperature} at room temperature. The experimentally measured electron and hole mobilities of GaN at such levels of doping are $\mu_n \sim 60 \frac{cm^2}{Vs}$  and $\mu_p \sim 3 \frac{cm^2}{Vs}$. These quantities can vary a bit depending on the quality of the sample and the growth conditions\cite{chin1994electron,jain2000iii,kozodoy2000heavy,nakamura1991highly}.  These mobilities result in broadenings of $\eta_c \sim 1.5 meV$ and $\eta_v \sim 3 meV$. In addition to these $\eta$ values, the device is simulated in two regimes; 1) with a reasonably small broadening of $\eta_c=\eta_v=1meV$ and 2) with a reasonably large broadening of $\eta_c=\eta_v=10meV$. This small and large broadening limits correspond to the largest and smallest mobilities respectively measured within an order of magnitude of the above specified doping level in different samples\cite{chin1994electron,jain2000iii,kozodoy2000heavy,nakamura1991highly}. As shown in Fig. \ref{IV_fig}, a good agreement is obtained between the experimental measurements and the simulated current-voltage characteristics from this work. On the other hand, ballistic transport models underestimate the tunnel current by several orders of magnitude.

\begin{figure}[H]
\centering
\includegraphics[width=180mm]{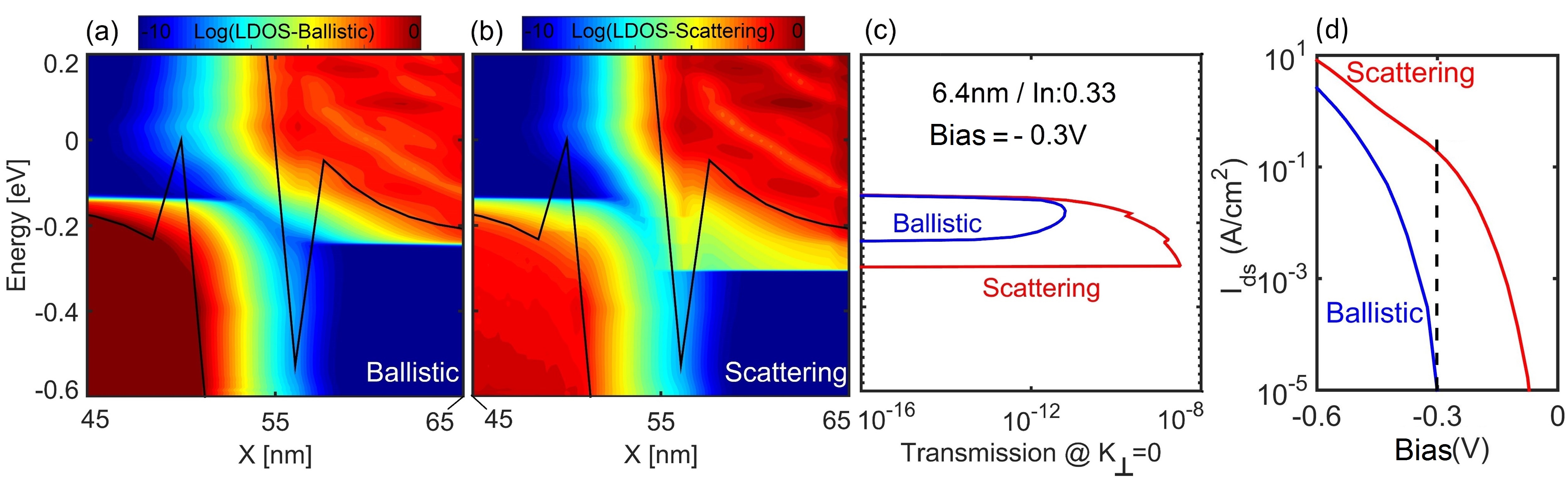}
\caption{a) Ballistic local density of states (LDOS) in the device measured in \cite{krishnamoorthy2010polarization} at -0.3V and K$_{\perp}$=0, the DOS is zoomed into the quantum well region. b) The corresponding scattering LDOS in the quantum well showing significantly larger DOS than the ballistic case. c) The ballistic v.s scattering transmission.  d) The ballistic versus scattering current. Ballistic transport significantly underestimate the LDOS and the transmission inside the quantum well resulting in a significantly less tunneling current as shown in Fig \ref{IV_fig}. Note that the quantum well of the device measured in \cite{krishnamoorthy2010polarization} is not deep enough to have resonant states. Resonant states, if present, provides a larger difference in the transport characteristics between the ballistic and scattering cases, as shown in Fig. \ref{LDOS_large_density_fig}.}
\label{LDOS_fig}
\end{figure}

\begin{figure}[H]
\centering
\includegraphics[width=140mm]{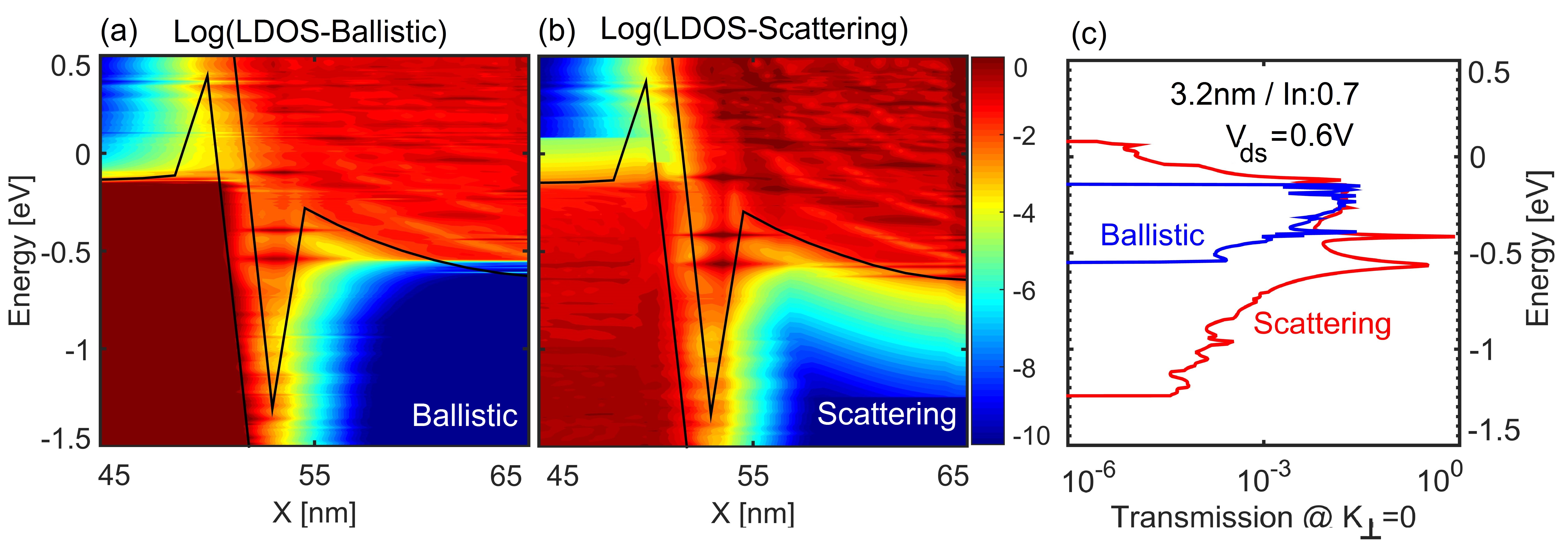}
\caption{Same as Fig. \ref{LDOS_fig} but a quantum well of 3.2nm and In$_{0.7}$Ga$_{0.3}$N is used instead. This case has a deeper quantum well that gives rise to resonant states which were absent in the measured device \cite{krishnamoorthy2010polarization}. Scattering is needed for proper broadening and filling of the resonant states.}
\label{LDOS_large_density_fig}
\end{figure}

\begin{figure}[H]
\centering
\includegraphics[width=60mm]{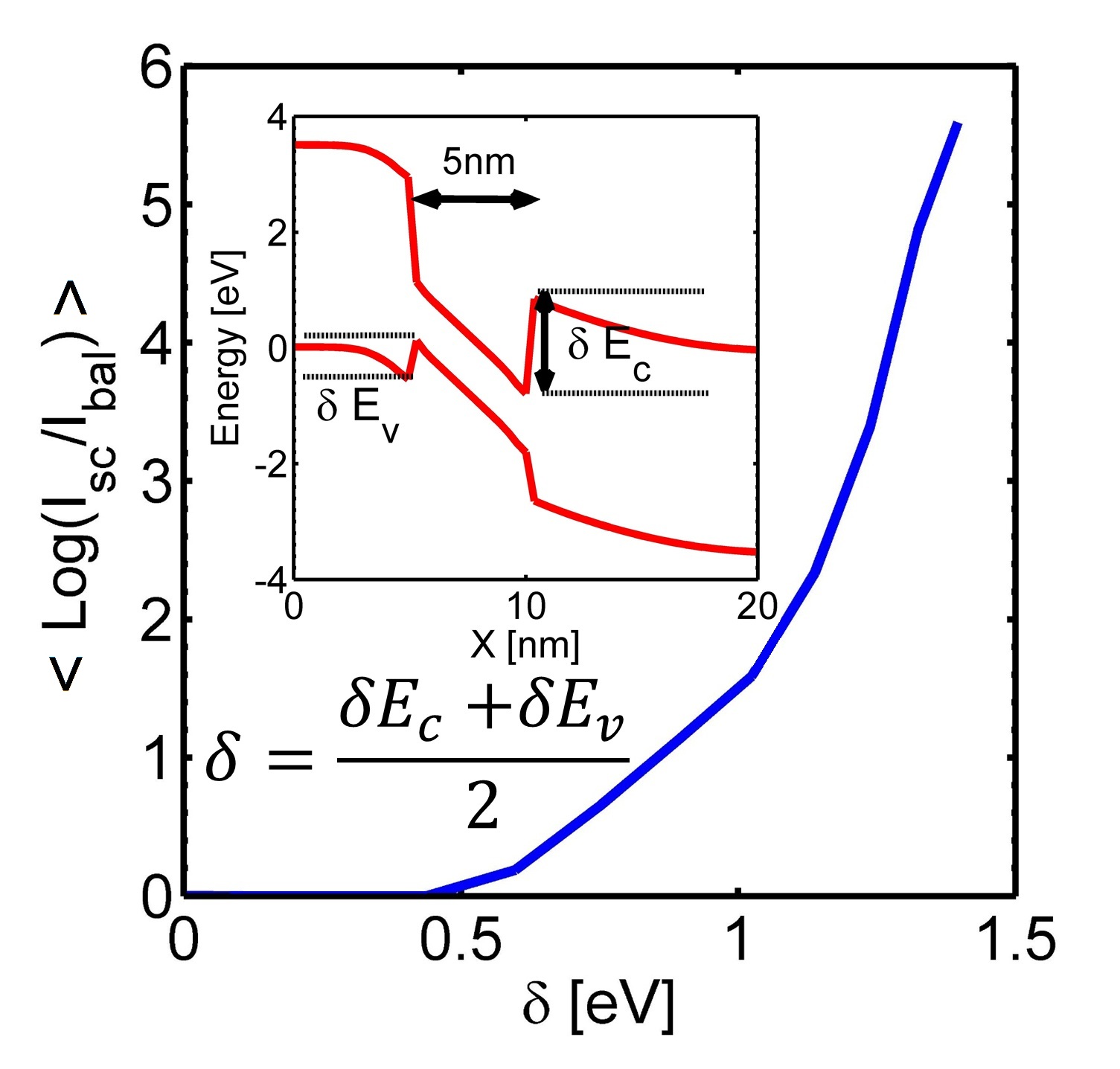}
\caption{Log of the ratio of scattering and ballistic currents (error) plotted against $\delta$ for a 5nm InGaN quantum well, where the average band offset $\delta=\frac{\delta E_c+\delta E_v}{2}$. $\delta$ is a measure of the amount of confined states influencing the transport. The error is averaged over $V_{ds}=0\rightarrow1V$. }
\label{Error_fig}
\end{figure}

 The reason behind the failure of ballistic model is that it does not account for the scattering induced broadening and filling of the confined states. Such effects are shown clearly in Fig. \ref{LDOS_fig} which compares the local density of states (LDOS) and transmission obained using ballistic and scattering models. The ballistic case doesn't show any density of states inside the quantum wells. As a result, the transmission and the current are underestimated. It is worth mentioning that the confined states in the device reported in \cite{krishnamoorthy2010polarization} do not have resonant states. If the quantum well is deep enough to have resonant states, the difference between ballistic and scattering models is even more pronounced as shown in Fig. \ref{LDOS_large_density_fig}. In order to show, the extent of such effect, the error between I-V of THJ from ballistic transport and scattering model is calculated as a function of band offsets. The error is defined as the order of magnitude difference between scattering and ballistic currents Log$\left(\frac{I_{sc}}{I_{bal.}}\right)$ averaged over a 1V bias voltage sweep. The error is calculated at different In concentrations of a 5nm InGaN quantum well, as shown in Fig. \ref{Error_fig}. The error is plotted against the average band offset $\delta=\frac{\delta E_c+\delta E_v}{2}$, this $\delta$ is a measure of the amount of confined states in the quantum wells at two sides of tunnel junction contributing to the tunneling current. Th error remains zero at small values of $\delta$, where no confined states exist in the quantum well. Hence, ballistic transport is acceptable for homo-junction tunneling devices\cite{luisier2010simulation,ilatikhameneh2015tunnel,ilatikhameneh2015can,ilatikhameneh2015scaling,ameen2016few,ilatikhameneh2016saving}. As $\delta$ increases, the error increases exponentially with $\delta$ and the scattering model becomes essential.  
  
\section{Conclusion}

In summary, we have developed a new quantum transport model for THJs. 
It is critical to use this model for THJs where strong scattering mechanisms instantly thermalize carriers in quantum wells adjacent to the tunnel junction, a situation that cannot be described well by the conventional ballistic models. The device characteristics obtained from the new model agrees fairly well with experimental measurements of Nitride THJs. Implementation of the new model in NEMO5 offers an efficient engineering tool that can be used to predict the performance and design THJs.


\section*{Appendix A: Strain and Polarization.}
Assuming the growth direction is  [001], the strain components $\epsilon_{ii}$ inside the quantum well are given by\cite{UBS}
\begin{equation}
\varepsilon_{xx}=\varepsilon_{yy} =\frac{a_{Substrate}-a_{well}}{a_{well}},
\label{eqn_App_1}
\end{equation}
\begin{equation}
\varepsilon_{zz} =\frac{-2 \nu\sigma_{xx}}{E},
\label{eqn_App_2}
\end{equation}
where $a$ is the lattice constant, $E$ is the Young's modulus and $\nu$ is Poisson ratio of the well material. All the material parameters for the In$_x$Ga$_{1-x}$N quantum well have been interpolated from InN and GaN. The effect of strain on the effective mass is small especially along the transport direction  c-axis\cite{dreyer2013effects}, and can be ignored.  
The strain slightly increases the band gap of the InGaN quantum well\cite{yan2009strain},
\begin{equation}
E_g=E_{g0}+(a_{cz}-D_{1}-D_3)\times\varepsilon_{zz}+(a_{ct}-D_{2}-D_4)\times(\varepsilon_{xx}+\varepsilon_{yy}),
\label{eqn_App_3}
\end{equation}
where $a_{cz}$, $a_{ct}$, $D_1$, $D_2$, $D_3$, and $D_4$ are deformation potentials for the In$_x$Ga$_{1-x}$N quantum well which are interpolated from InN and GaN deformation potentials\cite{yan2009strain}.
\begin{figure}[H]
\centering
\includegraphics[width=60mm]{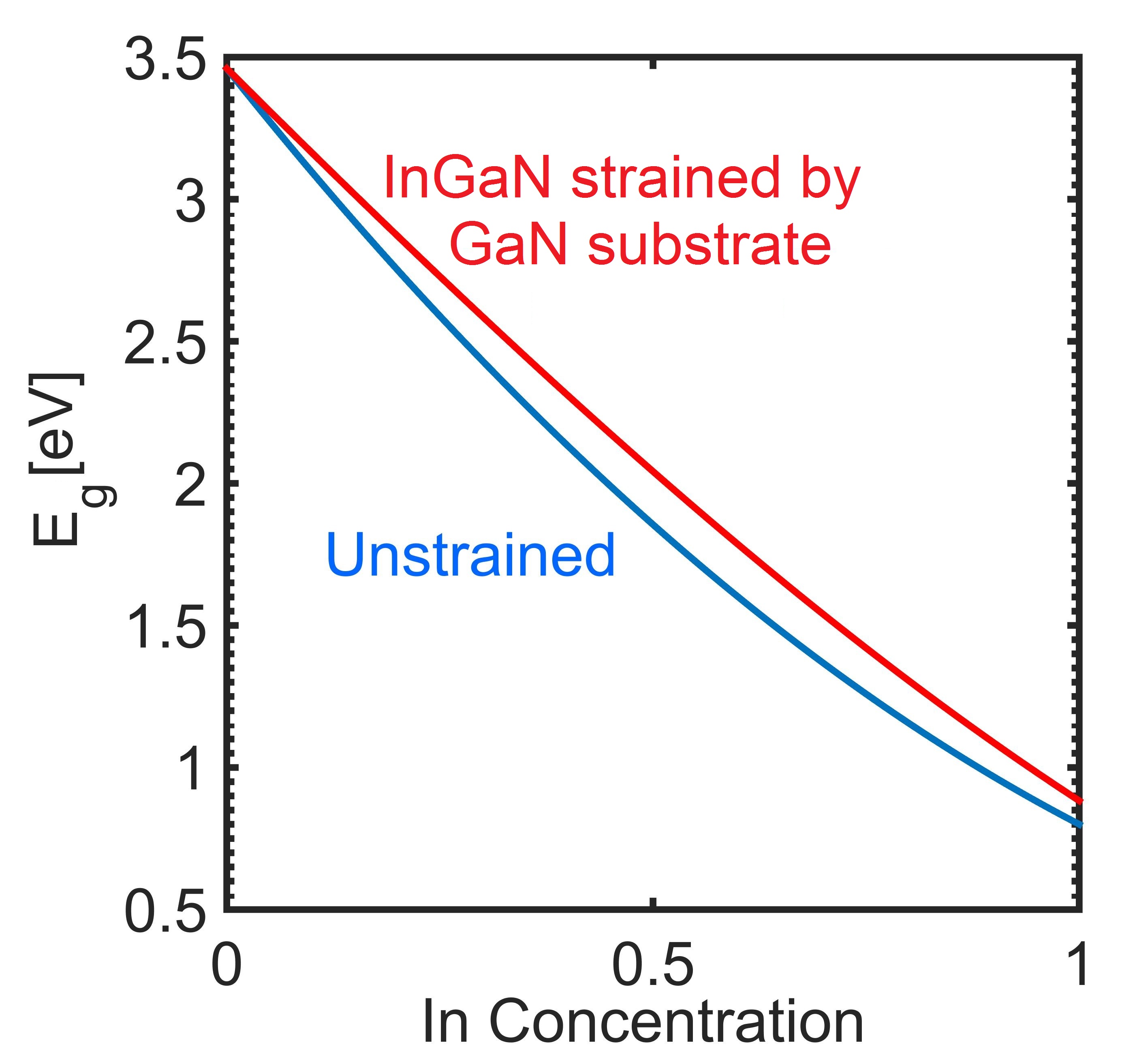}
\caption{Energy gap $E_g$ of In$_x$Ga$_{(1-x)}$N as a function of In concentration. Shown in blue is the  $E_g$ of the unstrained case, and in red, the $E_g$ of a quantum well strained by a GaN substrate.}
\label{Eg_fig}
\end{figure}
  
\begin{figure}[H]
\centering
\includegraphics[width=60mm]{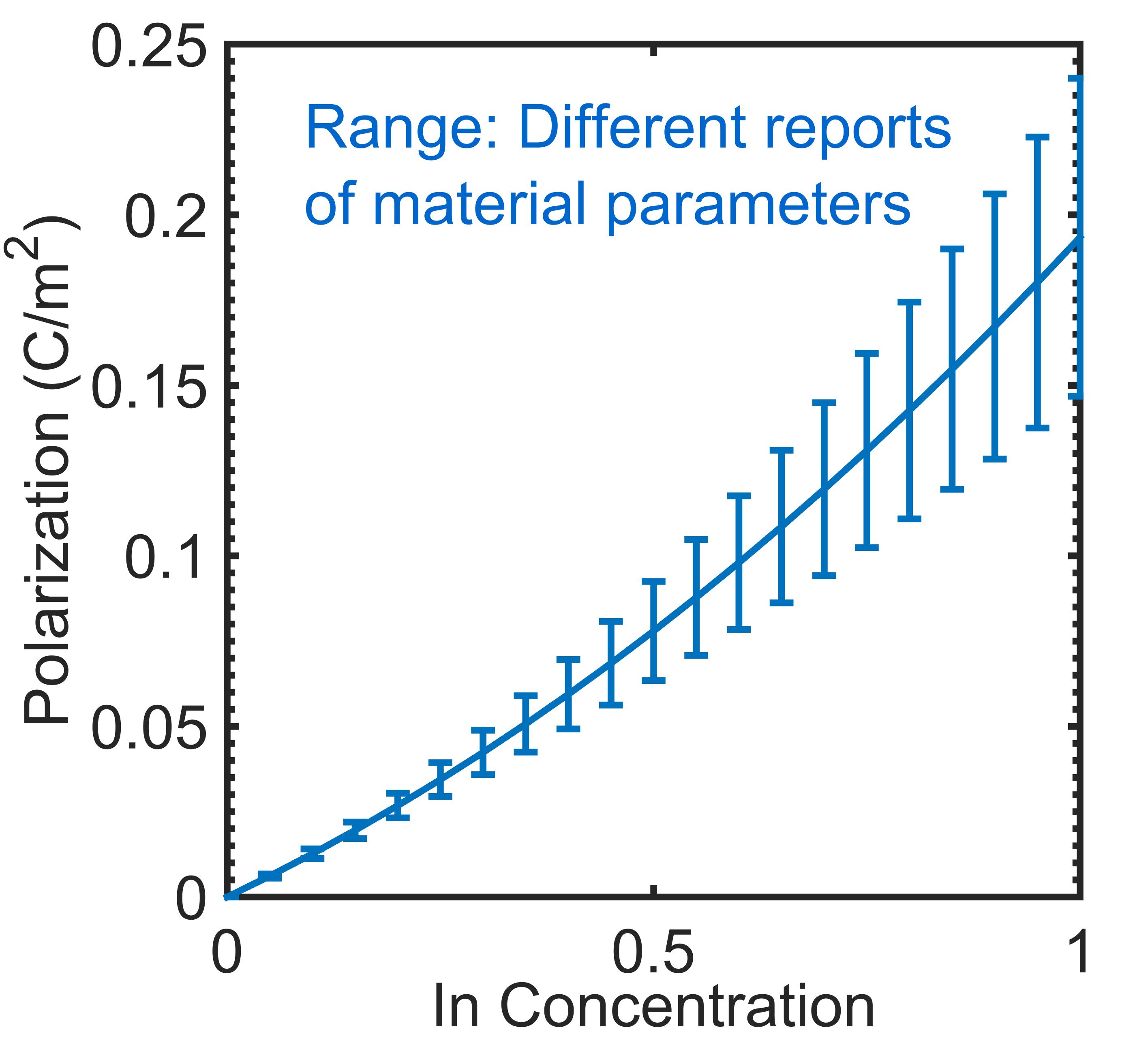}
\caption{Polarization of In$_x$Ga$_{(1-x)}$N quantum well on GaN substrate plotted against Indium concentration. The error bar indicates the range of polarization values obtained with different material parameters reported in literature. The average value represented by the solid-line has been used for the simulations in this work.}
\label{Polarization_fig}
\end{figure}

The piezoelectric polarization in the quantum well is given by
\begin{equation}
P_{piezo}=e_{33}\times \varepsilon_{zz}+2 e_{31}\times \varepsilon_{xx},
\label{eqn_App_3}
\end{equation}
where $e_{33}$ and $e_{31}$ are the interpolated piezoelectric coefficients for the InGaN quantum well.
There are many various reported values in literature from experimental measurements and first principle calculations of  Poisson's ratios and the piezoelectric coefficients for InN and GaN \cite{wagner2002properties,lepkowski2011poisson,piprek2007nitride,moram2007accurate,bernardini1997spontaneous,bernardini2001accurate}. The variations in these parameters result in a range of values for the polarization field of the InGaN quantum well especially at large In concentrations as shown in Fig. \ref{Polarization_fig}. In this work, we have used the average polarization value (shown by the solid line in Fig. \ref{Polarization_fig}) for the device simulations.

\bibliographystyle{ieeetr}
\bibliography{thesis_11_15_2016}

\section*{Acknowledgment}
This work was supported in part by the Center for Low Energy Systems Technology (LEAST), one of six centers of STARnet, a Semiconductor Research Corporation program sponsored by MARCO and DARPA. The use of nanoHUB.org computational resources operated by the Network for Computational Nanotechnology funded by the US National Science Foundation under Grant Nos. EEC-1227110, EEC-0228390, EEC-0634750, OCI-0438246, and OCI-0721680 is gratefully acknowledged. NEMO5 developments were critically supported by an NSF Peta-Apps award OCI-0749140 and by Intel Corp.

\end{document}